\documentclass[pra,twocolumn,superscriptaddress,showpacs,floatfix]{revtex4-1}
\usepackage{graphicx,amssymb}
\usepackage[fleqn]{amsmath}
\usepackage{amsfonts}
\usepackage{dcolumn}
\usepackage{bm}
\usepackage{braket}
\usepackage{mathtools}
\usepackage{multirow}

\begin{document}

\preprint{ }

\title{On the solution of a pairing problem in the continuum}

\author{A. Mercenne}
\affiliation{Grand Acc\'el\'erateur National d'Ions Lourds (GANIL), CEA/DSM - CNRS/IN2P3, BP 55027, F-14076 Caen Cedex, France}

\author{N. Michel}
\affiliation{NSCL/FRIB Laboratory, Michigan State University, East Lansing, Michigan 48824, USA}

\author{J. Dukelsky}
\affiliation{Instituto de Estructura de la Materia, CSIC, Serrano 123, E-28006 Madrid, Spain}

\author{M. P{\l}oszajczak}
\affiliation{Grand Acc\'el\'erateur National d'Ions Lourds (GANIL), CEA/DSM - CNRS/IN2P3, BP 55027, F-14076 Caen Cedex, France}

\date{\today}

\begin{abstract}
We present a generalized Richardson solution for fermions interacting with the pairing interaction in both discrete and continuum parts of the single particle (s.p.) spectrum. The pairing Hamiltonian is based on the rational Gaudin (RG) model which is formulated in the Berggren ensemble. We show that solutions of the generalized Richardson equations are exact in the two limiting situations: (i) in the pole approximation and (ii) in the s.p. continuum. If the s.p. spectrum contains both discrete and continuum parts, then the generalized Richardson equations provide accurate solutions for the Gamow Shell Model.

\end{abstract}

\maketitle

\section{Introduction}
\label{introduction}
{The pairing interaction is an important component of the effective nuclear interaction responsible for superfluid correlations and fluctuations from finite nuclei to neutron stars \cite{dean03_0}.
In most studies, approximate solutions of the pairing Hamiltonian are used to describe the structure of bound complex nuclei. Exact solutions of the pairing Hamiltonian for a constant pairing strength and a discrete set of s.p. levels are known since the seminal work of Richardson \cite{richardson63_1,richardson64_2}. By combining the Richardson exact solution with the integrable model proposed by Gaudin for quantum spin systems \cite{gaudin76_30}, it was possible to derive three classes of exactly solvable pairing models for fermions and bosons \cite{dukelsky01_31}. The Richardson or constant $g$ pairing Hamiltonian appears as a particular combination of the integrals of motion within the rational class of integrable models. However, more general exactly solvable pairing models could be derived from arbitrary combinations of the integrals of motion within each of the classes. In particular, a physically sound separable pairing Hamiltonian for heavy nuclei, derived from the hyperbolic family of Gaudin models, has been recently proposed in \cite{dukelsky11_12}. Moreover, rational Gaudin (RG) model has been extended to larger Lie algebras including the $SO(5)$ for $T=1$ isovector pairing \cite{dukelsky06_28} and the $SO(8)$ for $T=0,1$ spin-isospin pairing \cite{dukelsky07_34} allowing for the exact treatment of proton-neutron Hamiltonians.

There have been several attempts to formulate the exact solution of the pairing model in the continuum. Hasegawa and Kaneko studied effects of s.p. resonances (Gamow states) on pairing correlations \cite{hasegawa03_4}. Id Betan attempted to solve Richardson equations with the real or continuum-energy continuum \cite{idbetan12_7} but no proof was given that these equations yield exact solution of the pairing problem in the continuum. Such an exact solution can be obtained in the Gamow shell model (GSM) \cite{michel02_10,idbetan02_32,michel03_27,michel09_8} by exact diagonalization of the pairing Hamitonian. However, computer limitations restrict the calculations to systems with few active nucleons. Exactly solvable Hamiltonians could go beyond these limitations by reducing the complexity of an exact diagonalization to solving a small set of nonlinear equations. However, details of the mixing between the discrete s. p. levels with the continuum should be treated with extreme care in order to arrive to an exact solution.

In this paper, we formulate RG pairing model in the Berggren ensemble including s.p. bound states, resonances, and the non-resonant continuum. We show in Sect. \ref{generalized_RG_model} that the combination of the three ingredients yields a pairing model with a state-dependent pairing interaction that is not integrable in the general case. Hence, in Sect. \ref{sec2A} we derive a generalized Richardson solution for the RG model with the continuum which is exact in the pole approximation \cite{michel02_10,michel09_8} of this model and the two limiting cases: the discrete spectrum of real-energy s.p. levels and in the non-resonant s.p. continuum. By comparing with exact GSM solutions of this generalized RG pairing model (Sect. \ref{study_relative_error}), we discuss the salient features of generalized Richardson solutions in different sets of s.p. levels. Finally, in Sect. \ref{sec4} we summarize the main results of this work.
}

\section{The generalized rational Gaudin model}
\label{generalized_RG_model}
{
The constant pairing Hamiltonian derived from the rational RG model is given by:
	\begin{equation}
		H = \sum_{ \alpha }^{D} { \epsilon }_{ \alpha } { c }_{ \alpha }^{ \dagger } { c }_{ \alpha } + G \sum_{ \alpha , \beta }^{D} { c }_{ \alpha }^{ \dagger } { c }_{ \bar{\alpha} }^{ \dagger } { c }_{ \bar{\beta} } { c }_{ \beta }
		\label{pairing_hamiltonian}
	\end{equation}
where ${\epsilon }_{ \alpha }$ are the the energies of bound single-particle (s.p.) levels, and ${ G }$ is the pairing strength.
Operators ${ { c }_{ \alpha }^{ \dagger }({ c }_{ \alpha }) }$ stand for the particle creation (annihilation) operators, and ${ \alpha \equiv \{ a , { m }_{ \alpha } \} = \{ { n }_{ a } , { \ell }_{ a } , { j }_{ a } , { m }_{ \alpha } \} }$,  $ \bar{\alpha} = \{ a , { \bar{m} }_{ \alpha } \}$. ${ { c }_{ \bar{ \alpha } }^{ \dagger } }$ is defined as ${ { c }_{ \bar{\alpha} }^{ \dagger } = { (-) }^{ { j }_{ a } - { m }_{ \alpha } } { c }_{ \alpha , - { m }_{ \alpha } }^{ \dagger } }$. The degeneracy of a s.p. level $a$ is ${ \Omega }_{ a }=2  j _{ a }+1 $.

Let us define the operators
\begin{equation}
	{\hat n}_{a} = \sum_{ { m }_{ \alpha } = - { j }_{ a } }^{ { j }_{ a } } { c }_{ \alpha }^{ \dagger } { c }_{ \alpha }~~ ; ~~{ b }_{ a }^{ \dagger } = \sum_{ { m }_{ \alpha } > 0 } { c }_{ \alpha }^{ \dagger } { c }_{ \bar{\alpha}}^{ \dagger } = { ({ b }_{ a }) }^{ \dagger }
	\label{standard_commut_rel}
\end{equation}
which obey the SU(2) commutator algebra:
\begin{eqnarray}
\left[ { \hat n }_{ a } , { b }_{ a' }^{ \dagger } \right] &=& 2 { \delta }_{ aa' } { b }_{ a }^{ \dagger } \nonumber \\
\left[ { b }_{ a } , { b }_{ a' }^{ \dagger } \right] &=& 2 { \delta }_{ aa' } \left( \frac{ { \Omega }_{ a } }{ 4 } - \frac{ { \hat{n} }_{ a } }{ 2 } \right)
\label{reltwo}
\end{eqnarray}

The complete set of states of $N$ particles in $\cal{ N}$ s.p. states, spanned by the operators ${\hat n}_{a}$, ${ b }_{ a }$, ${ b }_{ a}^{ \dagger }$ is given by:
\begin{equation}
|n_1,n_2,\cdots,n_{\cal N},\nu\rangle=\frac{1}{\bar N}b_1^{\dagger n_1}b_2^{\dagger n_2}\cdots b_{\cal N}^{\dagger n_{\cal N}}|\nu\rangle
\label{normf}
\end{equation}
where $|\nu\rangle = |\nu_1,\nu_2\cdots \nu_{\cal N}\rangle$ is a state of the unpaired particles which satisfy:
\begin{equation}
b_{a}|\nu\rangle = 0~~ ; ~~{\hat n}_{a}|\nu\rangle = \nu_{a}|\nu\rangle
\end{equation}
$\bar N$ in Eq. (\ref{normf}) is the normalization constant and $\nu$ is the total number of the unpaired particles: $\nu = N - 2N_{\rm pair}$, where $N_{\rm pair}$ is the number of pairs.

The pairing Hamiltonian (\ref{pairing_hamiltonian}) expressed in the operators ${\hat n}_{a}$, ${ b }_{ a }$, ${ b }_{ a}^{ \dagger }$ reads:
\begin{equation}
	H = \sum_{ a }^{ {\cal N} } { \epsilon }_{ a } { \hat{ n } }_{ a } + G \sum_{ a , a' }^{ {\cal N} } { b }_{ a }^{ \dagger } { b }_{ a' }
	\label{pairing_hamiltonian_with_pairs_operator}
	\end{equation}
The exact solution of the pairing Hamiltonian (\ref{pairing_hamiltonian_with_pairs_operator}) with a discrete set of bound s.p. levels was found by Richardson \cite{richardson63_1,richardson64_2}. Later, it was shown that the model is quantum integrable by finding a complete set of integrals of motion in terms of which the Hamiltonian can be obtained as a linear combination \cite{cambiaggio97_29}.

For a given configuration of $\nu$ unpaired particles, the eigenvalue of the pairing Hamiltonian (\ref{pairing_hamiltonian_with_pairs_operator}) can be written as:
		\begin{equation}
			\tilde{ \mathcal{E} }^{(K)} = \sum_{ i = 1 }^{ { N }_{ pair } } { E }_{ i }^{(K)} + \sum_{ a = 1 }^{ {\cal N} } { \epsilon }_{ a } { \nu }_{ a } ~~~~~~~K=0,1,\dots, K_{\rm max}
		\label{total_energy}
	\end{equation}
where index $K$ enumerates the eigenstates in an ascending order of the excitation energy, and $K_{\rm max}+1$ is the total number of eigenstates.
In general, ${ \tilde{ \mathcal{E} } }^{(K)}$ can be complex and then ${ \mathcal{R} ( \tilde{ \mathcal{E} }^{(K)} ) = \mathcal{E} }^{(K)}$ is the energy, and ${ 2 \mathcal{I} ( \tilde{ \mathcal{E} }^{(K)} ) = \Gamma }^{(K)}$ is the corresponding width of the $K^{\rm th}$ eigenstate. For each eigenstate, the pair energies ${ { E }_{ i } }^{(K)}$ in (\ref{total_energy}) are obtained by solving ${ { N }_{ \text{pair} } }$ non-linear coupled equations:
	\begin{eqnarray}
		1  - 2G \sum_{ a }^{ {\cal N} } \frac{ { d }_{ a } }{ 2 { \epsilon }_{ a } - { E }_{ i }^{(K)} } + 2G \sum_{ j \neq i }^{ { N }_{ \text{pair} } } \frac{ 1 }{ { E }_{ i }^{(K)} - { E }_{ j }^{(K)} } = 0   \nonumber   \\~~~~(K=0,1,\dots, K_{\rm max})~~~~~
		\label{richardson_equations}
	\end{eqnarray}
with the initial conditions for pair energies which depend on the occupation of s.p. levels ${ \epsilon }_{ a }$ ($a=1,\dots ,{\cal N})$ in the limit of vanishing pairing strength. In the above equation: ${ { d }_{ a } = { \nu }_{ a } / 2 } - { \Omega }_{ a } / 4  $.
	
Generalization of the RG model to include the continuum part of a s.p. spectrum can be formulated in the Berggren s.p.  ensemble \cite{berggren68_11} which includes bound states $(b)$, resonances $(r)$, and non-resonant $(c)$ continuum states. In this representation, the pairing Hamiltonian is:
\begin{eqnarray}
	H &=& \sum_{ i \in b,r } { \epsilon }_{ i } { \hat{ n } }_{ i } + \sum_{ c } \int_{ { L }_{ c }^{ + } } { \epsilon }_{ { k }_{ c } } { \hat{ n } }_{ { k }_{ c } }  d { k }_{ c } \nonumber \\
	&+& G \sum_{ i , i' \in b,r } { b }_{ i }^{ \dagger } { b }_{ i' } + G \sum_{ c , c' } \int_{ { L }_{ c }^{ + } } { b }_{ { k }_{ c } }^{ \dagger } { b }_{ { k' }_{ c' } } d { k }_{ c } d { k' }_{ c' } \nonumber \\
	&+& G \sum_{ (i \in b,r) , c } \int_{ { L }_{ c }^{ + } } \left( { b }_{ { k }_{ c } }^{ \dagger } { b }_{ i } + { b }_{ i }^{ \dagger } { b }_{ { k }_{ c } } \right) d { k }_{ c }
		\label{new_pairing_hamiltonian}
	\end{eqnarray}
Sums over ${ c }, { c' }$ denote summations over different partial waves $(\ell, j)$ up to $({ \ell }_{ \rm max } , { j }_{ \rm max} )$. ${ { k }_{ c } }$ is related to the energy of a s.p. state $c$ in the non-resonant continuum: ${ \epsilon }_{ c } = { \hbar }^{ 2 } { k }_{ c }^{ 2 } / 2m$, and ${ m }$ is the particle mass.
The discrete sums run over the real energy bound s.p. states and the complex energy s.p. resonances enclosed in between the contour ${ { L }_c^{ + } }$ and the real $k$-axis. All resonances of the same quantum numbers $(\ell, j)$ have the same contour ${ { L }_{c(\ell, j)}^{ + } }$ in the complex $k$-plane. More about the complete Berggren s.p. ensemble and its application in many-body systems can be found in Ref. \cite{michel09_8}.

The pair creation (annihilation) operators satisfy the commutator relations (\ref{reltwo})
for the discrete (bound states and resonances) s.p. states, and
\begin{eqnarray}
\left[ { \hat n }_{ k_c } , { b }_{ k'_{c'} }^{ \dagger } \right] &=& 2 \delta (k_c - k'_c) { \delta }_{cc'}{ b }_{ k_c }^{ \dagger } \nonumber \\
\left[ { b }_{ k_c } , { b }_{ k'_{c'} }^{ \dagger } \right] &=& \delta (k_c - k'_c) { \delta }_{cc'}\frac{ { \Omega }_{ k_c } }{ 2 } - { \delta }_{ k_ck'_c } { \delta }_{cc'}{ \hat{n} }_{ k_c }
	\label{commutator_discrete_continuum}
\end{eqnarray}
for the non-resonant scattering s.p. states.

In all practical applications, the continuum has to be discretized. It is convenient to define new pair and number operators:
\begin{equation}
{ \hat{ \tilde{n} } }_{ q } = { w }_{ q } { \hat{ n } }_{ q }~~ ; ~~{ { \tilde{b} }_{ q }^{ \dagger } = \sqrt{ { w }_{ q } } { b }_{ q }^{ \dagger }  } = ({ \tilde{b} }_{ q })^{ \dagger }
\end{equation}
where index ${ q }$ runs over all bound, resonance and discretized scattering states in the Berggren basis.
${ { w }_{ q } }$ is a Gaussian weight of the integration procedure. For bound and resonance states, ${ { w }_{ q } } = 1$.

With this definition, all pair states are normalized to unity and treated on the equal footing. The new operators
${ \hat{ \tilde{n} }}_q, { \tilde{b} }_{ q }, { \tilde{b} }_{ q }^{ \dagger }$ satisfy the same SU(2) commutation relations as the operators ${\hat n}_i, { b }_{ i }, { { b }_{ i }^{ \dagger } }$ in discrete levels (Eq. (\ref{reltwo})):
\begin{eqnarray}
\left[ { \hat {\tilde n} }_{ q } , { {\tilde b} }_{ q' }^{ \dagger } \right] &=& 2 { \delta }_{ qq' } { {\tilde b} }_{ q }^{ \dagger } \nonumber \\
	\left[ { \tilde{b} }_{ q } , { \tilde{b} }_{ q' }^{ \dagger } \right] &=& 2 { \delta }_{ qq' } \left( \frac{ { \Omega }_{ q } }{ 4 } - \frac{ { \hat{ \tilde{n} } }_{ q } }{ 2 } \right)
	\label{commutator_new_operator}
\end{eqnarray}
The Hamiltonian of the generalized RG model (\ref{new_pairing_hamiltonian}) expressed in the operators ${ \hat{ \tilde{n} }}_q, { \tilde{b} }_{ q }, { \tilde{b} }_{ q }^{ \dagger }$  reads:
\begin{equation}
		H = \sum_{ q }^{ {\cal N} } { \epsilon }_{ q } { \hat{ \tilde{n} } }_{ q } + \sum_{ q , q' }^{ {\cal N} } { G }_{ qq' } { \tilde{b} }_{ q' }^{ \dagger } { \tilde{b} }_{ q } ~~;~~  { { G }_{ qq' } = \sqrt{ { w }_{ q } } \sqrt{ { w }_{ q' } } } G
		\label{new_hamiltonian_discretized}
	\end{equation}
where ${\cal N}$ is the total number of bound, resonance and discretized continuum s.p. states.
In general, pairing models with the state-dependent pairing interaction are not integrable with the exception of the hyperbolic model \cite{dukelsky11_12,rombouts10_35} where the Gaussian weights $w_q$  should be a linear function of the s. p. energies $\epsilon_q$ in order for the system to be exactly solvable. One has then to look for reliable approximations to the Hamiltonian (\ref{new_hamiltonian_discretized}) or to the commutation relations (\ref{commutator_discrete_continuum}) for the non-resonant scattering states, which break the SU(2) commutator algebra, that could lead to an ansatz for an exact eigenstate.
	
It is important to note that if we want to diagonalize the Hamiltonian (Eq.(\ref{new_hamiltonian_discretized})) we have to be careful applying the new normalized operators ${ { \hat{ \tilde{ n } } }_{ q } }$ and ${ { \tilde{ b } }_{ q }^{ \dagger }, { \tilde{ b } }_{ q } }$.
As the Hamiltonian in Eq.(\ref{new_pairing_hamiltonian}) is expressed in a certain Slater determinant basis, the contour discretization leads not only to new normalized operators but also to new normalized Slater determinants, so that the action of ${ { \hat{ \tilde{ n } } }_{ q }, { \tilde{ b } }_{ q }^{ \dagger } }$ and ${ { \tilde{ b } }_{ q } }$ on it is defined as in the discrete case.

	\subsection{An approximate solution for the rational Gaudin model with the continuum }
	\label{sec2A}
	{
An approximate solution for the generalized rational pairing model (\ref{new_hamiltonian_discretized}) can be found by replacing the Kronecker delta by the Dirac delta in the commutator (\ref{commutator_discrete_continuum}) for states in the non-resonant continuum:
\begin{equation}
	\left[ { b }_{ k_c } , { b }_{ k'_{c'} }^{ \dagger } \right] = 2 \delta (k_c - k'_c){ \delta }_{cc'} \left( \frac{ { \Omega }_{ k_c } }{ 4 } - \frac{ { \hat{n} }_{ k_c } }{ 2 } \right) \ .
		\label{modified_commutator}
\end{equation}
 With this change, the pair operators ${ { \tilde{b} }_{ q }^{ \dagger } ({ \tilde{b} }_{ q }) }$  for bound, resonance and discretized scattering states satisfy:
\begin{eqnarray}
\left[ {\hat { \tilde n }}_{ q } , {\tilde { b }}_{ q' }^{ \dagger } \right] &=& 2 { \delta }_{ qq' } {\tilde { b }}_{ q }^{ \dagger }\nonumber \\
	\left[ { \tilde{b} }_{ q } , { \tilde{b} }_{ q' }^{ \dagger } \right] &=& 2 { \delta }_{ qq' } \left( \frac{ { \Omega }_{ q } }{ 4 } - \frac{ { \hat{ \tilde{n} } }_{ q } }{ 2 { w }_{ q } } \right)
		\label{modified_commutator_new_operator}
\end{eqnarray}

The transformation presented in Eq.(\ref{modified_commutator_new_operator}) is mathematically undefined.
Due to this choice, we cannot have a proper definition of these new operators and a direct diagonalization of the Hamiltonian Eq.(\ref{new_hamiltonian_discretized}) with the deformed operators (\ref{modified_commutator_new_operator}) is not possible.
In the following, to perform the derivation of our Richardson equations, we supposed that these deformed operators act like those in Eq.(\ref{standard_commut_rel}).

Let us now derive the eigenvalues of the pairing Hamiltonian (\ref{new_hamiltonian_discretized}) in this approximation.
Similarly as in the Richardson solution, each eigenstate $K$ ($K=1,2,\dots,K_{\rm max})$ is written as a product of the pair states:
\begin{equation}
	|{ \Psi }_{\rm norm}\rangle = \prod_{ \eta = 1 }^{ N_{\rm pair} } { B }_{ { \eta ; {\rm norm} } }^{ \dagger} | \nu \rangle
	\end{equation}
It is tacitly understood that each eigenstate, its pair creation (annihilation) operators and corresponding pair energies carry an index $K$ of the state. In the above expression, the pair operators are given by:
	\begin{equation}
	{ B }_{ { \eta ; {\rm norm} } }^ { \dagger} = { c }_{ { \eta } } G \sum_{ q }^{ N_{\rm pair} } \frac{ { \tilde{b} }_{ q }^{ \dagger } \sqrt{ { w }_{ q } } }{ 2 { \epsilon }_{ q } - { E }_{ { \eta } } }
\end{equation}
where ${ E }_{ { \eta } }$ are the pair energies in the eigenstate $K$.
The normalization constants ${ c }_{ \eta }$ are determined by solving:
\begin{equation}
	\frac{ 1 }{ { ( { c }_{ \eta } G ) }^{ 2 } } = \frac{ 1 }{ {( { { C }_{ \eta }} )}^{ 2 } } = \sum_{ q }^{ N_{\rm pair} } \frac{ { w }_{ q } }{ { (2 { \epsilon }_{ q } - { E }_{ \eta }) }^{ 2 } }
	\label{normalization_constant}
\end{equation}
In order to simplify the notation, it is convenient to define: ${ { B }_{ { \eta } }^{ \dagger} = { B }_{ { \eta ; {\rm norm} } }^ { \dagger} / { C }_{ \eta } }$, so that
\begin{equation}
	| { \Psi }_{\rm norm} \rangle = \prod_{ \eta = 1 }^{ N_{\rm pair} } { C }_{ \eta } { B }_{ { \eta } }^{ \dagger} |\nu \rangle = { C } | { \Psi } \rangle
	\label{}
\end{equation}
where $${ { C } = \prod_{ \eta = 1 }^{ N_{\rm pair} } { C }_{ { \eta } } }~~ {\rm and}~~ { | { \Psi } \rangle = \prod_{ \eta = 1 }^{ N_{\rm pair} } { B }_{ { \eta } }^{ \dagger } |\nu \rangle }~.$$
The operators ${ \hat{ \tilde{n} } }, { B }_{ { \eta } }$ and $B_0$:
\begin{equation}
{ { B }_{ 0 }^{ \dagger } = \sum_{ q }^{ {\cal N} } { \tilde{b} }_{ q }^{ \dagger } \sqrt{ { w }_{ q } } }
\end{equation}
satisfy the commutator relations:
\begin{eqnarray}
	 &&\left[ { \hat{ \tilde{n} } }_{ q } , { B }_{ { \eta } }^{ \dagger } \right] = \frac{ 2 { \tilde{b} }_{ q }^{ \dagger } \sqrt{ { w }_{ q } } }{ 2 { \epsilon }_{ q } - { E }_{ { \eta } } } \nonumber \\
	 &&\left[ { \hat{ \tilde{n} } }_{ q } , { B }_{ 0 }^{ \dagger } \right] = 2 \sqrt{ { w }_{ q } } { \tilde{b} }_{ q }^{ \dagger } \nonumber \\
	 &&\left[ { B }_{ { \eta } } , { B }_{ { \eta' } }^{ \dagger } \right] = \sum_{ q }^{ \cal{N} } \frac{ 2 { w }_{ q } ({ \Omega }_{ q } / 4 - { \nu }_{ q } / 2) }{ (2 { \epsilon }_{ q } - { E }_{ { \eta } }) (2 { \epsilon }_{ q } - { E }_{ { \eta' } }) } \nonumber \\
&&	\left[ { B }_{ 0 } , { B }_{ { \eta } }^{ \dagger } \right] = \sum_{ q }^{ {\cal N} } \frac{ { w }_{ q } { \Omega }_{ q } / 2 - { \hat{ \tilde{n} } }_{ q } }{ 2 { \epsilon }_{ q } - { E }_{ { \eta } } }  \nonumber \\
&& \left[ { B }_{ 0 }^{ \dagger } , { B }_{ { \eta } }^{ \dagger } \right] = 0 	
	\label{relation_B_O_B_j}
\end{eqnarray}
which can be derived from the commutation relations for operators ${\tilde { \hat n }}_{ q }, { { \tilde{b} }_{ q }^{ \dagger }, { \tilde{b} }_{ q } }$ (Eq. (\ref{modified_commutator_new_operator})).

The Hamiltonian of the generalized RG model (\ref{new_hamiltonian_discretized}) expressed in these operators reads:
\begin{equation}
	H = \sum_{ q }^{ {\cal N} } { \epsilon }_{ q } { \hat{ \tilde{n} } }_{ q } + G { B }_{ 0 }^{ \dagger } { B }_{ 0 } 	\label{discretized_hamiltonian}
\end{equation}
The pair energies defining each eigenstate correspond to a particular solution of the set of $N_{pair}$ nonlinear coupled Richardson type equations:

\begin{eqnarray}
	1 -2G \sum_{ q }^{ {\cal N} } \frac{  { w }_{ q } \left( { \nu }_{ q } / 2 - { \Omega }_{ q } / 4 \right)}{ 2 { \epsilon }_{ q } - { E }^{(K)}_{ { \eta } } } + 2G \sum_{ \eta' = 1 ; \neq \eta }^{ N_{\rm pair} } \frac{ 1 }{ { E }^{(K)}_{ { \eta } } - { E }^{(K)}_{ { \eta' } } } = 0
	\nonumber   \\~~~~(K=0,1,\dots, K_{\rm max})~~~~~~~~~
	\label{richardson_equations}
\end{eqnarray}
  The first sum in these generalized Richardson equations can be split on to the separate terms from the resonant states and the discretized scattering states.

 In the continuum limit, the generalized Richardson equations become:
 \begin{eqnarray}
	1 &-& 2G \sum_{ i \in b,r }^{ \cal N } \frac{ { d }_{ i } }{ 2 { \epsilon }_{ i } - { E }^{(K)}_{ { \eta } } } \nonumber \\
	&-& 2G \sum_{c}^{ { \ell }_{ \rm max} , { j }_{ \rm max} } \int_{ { L }_{ c }^{ + } } \frac{ { d }_{ { k }_{ c } } }{ { \hbar }^{ 2 } { k }_{ c }^{ 2 } / m - { E }^{(K)}_{ { \eta } } } d { k }_{ c }\nonumber \\
	&+& 2G \sum_{ \eta' = 1 ; \neq \nu }^{ N_{\rm pair} } \frac{ 1 }{ { E }^{(K)}_{ { \eta } } - { E }^{(K)}_{ { \eta' } } } = 0
	\nonumber   \\&&~~~~~~~~~~~~~~~~~~~~~~~~(K=0,1,\dots, K_{\rm max})~~
\end{eqnarray}
where ${ { d }_{ i } = { \nu }_{ i } / 2 - { \Omega }_{ i } / 4 }$ and similarly for $d_{{ k }_{ c } }$.

The generalized Richardson equation (\ref{richardson_equations}) is an approximate solution for the RG model with the continuum obtained by replacing the exact commutator relations (\ref{commutator_discrete_continuum}) by the approximate ones (\ref{modified_commutator}). In certain limiting situations, however, this solution is exact.  For a discrete set of bound s.p. levels, all weights $w_q$ are equal 1 and, hence, Eq. (\ref{richardson_equations}) reduces to an exact solution for the RG model \cite{richardson63_1,richardson64_2}. By the same argument, Eq. (\ref{richardson_equations}) provides an exact solution for the pairing model with the continuum in the pole approximation, {\it i.e.} neglecting the non-resonant continuum states.
Eq. (\ref{richardson_equations}) is also exact if the Berggren ensemble contains only discretized states of the non-resonant continuum because in this case one may take the same weights $w_q \equiv w$ for all continuum states $q$ and renormalize the pairing strength ${ G' = G w }$ accordingly. In this particular case, the third sum in Eq. (\ref{richardson_equations}) goes to 0 and one obtains:
\begin{eqnarray}
	1 - 2G \sum_{c}^{ { \ell }_{ \rm max} , { j }_{ \rm max} } \int_{} \frac{ { d }_{ k_c } }{ 2 { \epsilon }_{ k_c } - { E }_{ \eta} } dk_c = 0
	\nonumber   \\~~~~~~~~~~~~~~~~~~~~~~~~~~~(K=0,1,\dots, K_{\rm max})
	\label{final_continuum_richardson_equations}
\end{eqnarray}

	}
}

\section{Numerical solution of the rational Gaudin model with the continuum}
\label{resolution_richardson_equations}
{
	The numerical solution of generalized Richardson equations (\ref{richardson_equations}) is plagued by divergencies taking place when two or more pair energies coincide with twice a s.p. energy.
	In the weak coupling limit ($G\rightarrow 0$), the standard way to approach this problem is to start with an educated guess for pair energies $E_i$ and then evolve them by iteratively solving the generalized Richardson equations for increasing values of $G$.
	At each step, the solution for pair energies is updated with the Newton-Raphson method using the solution of the previous step as the new starting point \cite{rombouts04_23}.
	
	This initial guess is determined by solving the generalized Richardson equations in the limit $G\rightarrow 0$.
	The general expression for pair energies ${ { E }_{ i } }$ in this limit is:
	\begin{equation}
		\lim_{G \to 0} { E }_{ i } = 2 { \epsilon }_{ q } \;\;\; \text{ with } \;\;\; i = 1,\cdots, { N }_{ \text{ pair } } \;\; \text{ and } \;\; q = 1,\cdots,\mathcal{N}
		\label{boundary_conditions}
	\end{equation}
	The analytical determination of pair energies becomes difficult if many pairs occupy the same s.p. level
	${ q }$. In a general case of ${ { N }_{ \text{ pair } } }$ pairs occupying the same s.p. state of energy ${ { \epsilon }_{ q } }$,
	the starting pair energies ${ { E }_{ i } }$  are found by solving the set of ${ { N }_{ \text{ pair } } }$ coupled equations:
	\begin{equation}
		1 - \frac{ 2 G { d }_{ q } }{ 2 { \epsilon }_{ q } - { E }_{ i } } + 2G \sum_{j \neq i}^{N_{\rm pair}} \frac{ 1 }{ { E }_{ i } - { E }_{ j } } = 0  \;\;\;	\;\;\; i = 1,\cdots, { N }_{ \text{ pair } }
		\label{rich_eq_weak_coupling_limit}
	\end{equation}
	Notice that the non-resonant continuum states in the weak coupling limit ${ G << 1 }$ are not occupied and, hence, the corresponding terms in  generalized Richardson equations are absent in this limit.
	
	It is possible to write the analytic solution of Eq. (\ref{rich_eq_weak_coupling_limit}) for one or two pairs of particles on the same level ${ q }$. If a degeneracy of the s.p. level $q$ is $\Omega_q=2$, {\em i.e.} at most one pair of particles can occupy this level, the solution of Eq. (\ref{rich_eq_weak_coupling_limit}) is:
	\begin{equation}
		{ E }_{ i } = 2 { \epsilon }_{ q } - 2 G { d }_{ q }
		\label{one_pair_on_q}
	\end{equation}
	For higher degeneracy of s.p. states $q$ (${ { \Omega }_{ q } \geq 4 }$), the analytical solution of Eq. (\ref{rich_eq_weak_coupling_limit}) for two pairs of particles is:
	\begin{align}
		& { E }_{ i } = 2 { \epsilon }_{ q } - G ({ d }_{ q } + 1) + i  G \sqrt{ 2 { d }_{ q } + 1 } \nonumber \\	
		& { E }_{ i' } = 2 { \epsilon }_{ q } - G ({ d }_{ q } + 1) - i  G \sqrt{ 2 { d }_{ q } + 1 } 	
		\label{two_pairs_on_q}
	\end{align}
	For three pairs occupying the same level ${ q }$ at ${ G << 1 }$, we can use a combination of the solutions (\ref{one_pair_on_q}) and (\ref{two_pairs_on_q}), {\em i.e.} one pair is initiated with Eq. (\ref{one_pair_on_q}) while the two others are initiated with Eq. (\ref{two_pairs_on_q}).
	
	It is interesting to notice that if two pairs at ${ G \rightarrow 0 }$ occupy the same s.p. state ${ q }$, then their energies are complex conjugate.
	If the s.p. spectrum is real then this symmetry of the pair energies at ${ G \rightarrow 0 }$ is preserved by the iterative procedure of solving the generalized Richardson equations for any ${ G }$.
	This special symmetry of pair energies in the weak coupling limit is broken for finite $G$ if the non-resonant continuum states are included in the basis.
	Indeed, continuum states are absent in Eq. (\ref{rich_eq_weak_coupling_limit}) but become occupied for finite values of the pairing strength ${ G }$ and hence, the initial symmetry of pair energies is broken in the course of solving the generalized Richardson equations.
	
	For systems with an odd number particles and/or broken pairs, the generalized seniority $\nu$ is different from 0. Each configuration is defined by the set seniorities ${ { \nu }_{ q } }$  giving the information of how many of unpaired particles occupy the level ${ q }$.  The same  Eqs. (\ref{one_pair_on_q}), (\ref{two_pairs_on_q}) are then used
  to obtain an initial guess for the pair energies and initiate the iterative procedure.

	Numerical solutions of the generalized Richardson equations exhibit singularities also for finite ${ G }$ \cite{richardson65_20}.
	Formally, they cancel out and the total energy (the sum of pair energies) is always a continuous function of ${ G }$. However, these singularities generate instabilities in numerical applications which are hard to deal with. Those which occur at specific values of the pairing strength ${ { G }_{ c } }$, are seen in the convergence of different pair energies to the same energy ${ 2 { \epsilon }_{ q } }$. Close to this critical point, the derivative of pair energies with respect to ${ G }$ becomes very large making the Newton-Raphson method unstable.
	
	The practical solution of this problem has been proposed by Richardson in the case of doubly degenerate levels \cite{richardson66_21}. In this case, two pair energies ${ { E }_{ \lambda } }$ and ${ { E }_{ \lambda' } }$ converge to the same energy ${ 2 { \epsilon }_{ q } }$, thus it is convenient to use a new set of variables:
	\begin{align}
		& { \lambda }_{ + } = 4 { \epsilon }_{ q } - { E }_{ \lambda } - { E }_{ \lambda' } \nonumber \\
		& { \lambda }_{ - } = { \left( { E }_{ \lambda } - { E }_{ \lambda' } \right) }^{ 2 }
		\label{new_variables}
	\end{align}
 for ${ G\simeq{ G }_{ c } }$. The particularity of these new variables is that their derivative with respect to $G$ does not diverge at ${ G={ G }_{ c } }$. Thus, it is possible to perform a polynomial fit of ${ \lambda }_{ + }(G)$ and ${ \lambda }_{ - }(G)$ in the vicinity of ${ { G }_{ c }  }$, and extrapolate the pair energies  ${ { E }_{ \lambda } }$ and ${ { E }_{ \lambda' } }$ across ${ { G }_{ c } }$.	
				
	As a first test  of the approximate rational RG model with the continuum we compare it with an exact Gamow shell model diagonalization of the pairing Hamiltonian (\ref{new_pairing_hamiltonian}).  We discretize the contour ${ { L }_c^{ + } }$ using the Gauss-Legendre quadrature method and build the s.p. spectrum which is used both in the generalized Richardson equations (\ref{richardson_equations}) and in the GSM.

\subsubsection{Numerical solution of pairing Hamiltonian in the GSM}
	\label{resolution_gsm_equations}
	{
	
	Exact solutions of the constant pairing Hamiltonian (\ref{new_pairing_hamiltonian}) are obtained by diagonalizing the Hamiltonian matrix using the Davidson method. This matrix is sparse with only $\sim$0.4\% of non-zero matrix elements. The calculation of eigenvalues in this case is efficient because  matrix-vector multiplications are fast and the storage of a matrix can be optimized.
	
	}
	
		\subsubsection{Calculation of the pairing gap}
		{
		A useful measure of pairing correlations in a given eigensate ${ \ket{ { \Psi }^{ (K) } } }$ is the canonical pairing gap:	
		\begin{equation}
			{ \Delta }^{ (K) } = G \sum_{q}^{ \mathcal{N} } \sqrt{ { n }_{ q }^{ (K) } (1 - { n }_{ q }^{ (K) }) }
			\label{gap_expression}
		\end{equation}	
		where the sum runs over s.p. states, and ${ { n }_{ q }^{ (K) } }$ is the occupation probability of the state ${ q }$. We note that the canonical gap coincides with the BCS gap in the BCS approximation where the occupation probability ${ { n }_{ q }^{ (K) }= |v_q|^2 }$. The determination of the occupation probability ${ { n }_{ q }^{ (K) } }$ can be done exactly through the diagonalization of GSM Hamiltonian. 		
		Let us write the eigenstate ${ \ket{ { \Psi }^{ (K) } } }$ of a pairing Hamiltonian (\ref{pairing_hamiltonian_with_pairs_operator}) as an expansion in a basis of Slater determinants ${ \ket{ { \Phi }_{ \alpha } } }$ :
		\begin{equation}
			\ket{ { \Psi }^{ (K) } } = \sum_{ \alpha } { C }_{ \alpha }^{ (K) } \ket{ { \Phi }_{ \alpha } }	
			\label{}
		\end{equation}
		The expectation value of the particle number operator $\hat{ N }$ is:
		\begin{align}
			N = \bra{ { \Psi }^{ (K) } } \hat{ N } \ket{ { \Psi }^{ (K) } } & = \sum_{\alpha , \alpha'} { C }_{ \alpha }^{ (K) } { C }_{ \alpha' }^{ (K) } \bra{ { \Phi }_{ \alpha } } \hat{ N } \ket{ { \Phi }_{ \alpha' } } \nonumber \\
			& = \sum_{q} 2 { n }_{ q }^{ (K) }
			\label{N_nk}
		\end{align}
		Hence, the occupation probability can be determined numerically as:
		\begin{equation}
			{ n }_{ q }^{ (K) } = \sum_{\alpha} g(\alpha,q;K) \; { ( { C }_{ \alpha }^{ (K) } ) }^{ 2 }
			\label{nk_exact}
		\end{equation}
		where ${ g (\alpha,q;K) }$ is equal to 1 or 0 depending on whether the s.p. state ${ q }$ is occupied or unoccupied in the Slater determinant ${ \alpha }$ of an eigenstate ${ K }$.
		
		In the generalized Richardson equations the s.p. occupation probabilities in an eigenstate ${ K }$ are determined by means of the Hellmann-Feynman theorem \cite{richardson65_20,dukelsky02_37}:
		\begin{equation}
			{ n }_{ q }^{ (K) } = \frac{ \partial { \tilde{ \mathcal{E} } }^{ (K) } }{ \partial { \epsilon }_{ q } } \ ,
			\label{nk_richardson}
		\end{equation}
		where ${ { \tilde{ \mathcal{E} } }^{ (K) } }$ is the total energy (\ref{total_energy}) of the eigenstate ${ K }$.		
		
}

\subsection
{Comparison between solutions of GSM and generalized Richardson equations}
\label{study_relative_error}
	{

	\subsubsection{Bound single particle states}
		{
		In this subsection, we compare results obtained by solving the generalized Richardson equations (\ref{richardson_equations}) for fermions with the exact GSM results for a spectrum of well-bound s.p. levels: ${ { \epsilon }_{ i } = \{ -5 , -4 , -3 , -2 , -1 \} }$ MeV.
		\begin{table*}[t]
	\begin{ruledtabular}
		\begin{tabular}{ccccccc}
			G (MeV) & Nb of pairs & 9pts & 15pts & 21pts & 30pts & 45pts \\
			\hline
			\multirow{5}{*}{-0.01} & 2 pairs & ${ { 7.2138 \text{e} }^{ -13 } }$ & ${ { 7.7698 \text{e} }^{ -13 } }$ & ${ { 8.0458 \text{e} }^{ -13 } }$ & ${ { 8.2666 \text{e} }^{ -13 } }$ & ${ { 8.4342 \text{e} }^{ -13 } }$ \\
			 & 3 pairs & ${ { 2.3513 \text{e} }^{ -12 } }$ & ${ { 2.5342 \text{e} }^{ -12 } }$ & ${ { 2.6134 \text{e} }^{ -12 } }$ & ${ { 2.6800 \text{e} }^{ -12 } }$ & ${ { 2.7341 \text{e} }^{ -12 } }$ \\
			 & 4 pairs & ${ { 6.7547 \text{e} }^{ -12 } }$ & ${ { 7.2574 \text{e} }^{ -12 } }$ & ${ { 7.4856 \text{e} }^{ -12 } }$ & ${ { 7.6711 \text{e} }^{ -12 } }$ & ${ { 7.8102 \text{e} }^{ -12 } }$ \\
			 & 5 pairs & ${ { 2.5847 \text{e} }^{ -11 } }$ & ${ { 2.7248 \text{e} }^{ -11 } }$ & ${ { 2.8080 \text{e} }^{ -11 } }$ & ${ { 2.8722 \text{e} }^{ -11 } }$ & ${ { 2.9235 \text{e} }^{ -11 } }$ \\
			\hline
			\multirow{5}{*}{-0.3} & 2 pairs & ${ { 1.2957 \text{e} }^{ -6 } }$ & ${ { 1.3994 \text{e} }^{ -6 } }$ & ${ { 1.4467 \text{e} }^{ -6 } }$ & ${ { 1.4840 \text{e} }^{ -6 } }$ & ${ { 1.5142 \text{e} }^{ -6 } }$ \\
			& 3 pairs & ${ { 3.7227 \text{e} }^{ -6 } }$ & ${ { 4.0202 \text{e} }^{ -6 } }$ & ${ { 4.1560 \text{e} }^{ -6 } }$ & ${ { 4.2630 \text{e} }^{ -6 } }$ & ${ { 4.3496 \text{e} }^{ -6 } }$ \\
			& 4 pairs & ${ { 9.3104 \text{e} }^{ -6 } }$ & ${ { 1.0038 \text{e} }^{ -5 } }$ & ${ { 1.0375 \text{e} }^{ -5 } }$ & ${ { 1.0642 \text{e} }^{ -5 } }$ & ${ { 1.0858 \text{e} }^{ -5 } }$ \\
			& 5 pairs & ${ { 2.8099 \text{e} }^{ -5 } }$ & ${ { 2.9908 \text{e} }^{ -5 } }$ & ${ { 3.0917 \text{e} }^{ -5 } }$ & ${ { 3.1709 \text{e} }^{ -5 } }$ & ${ { 3.2352 \text{e} }^{ -5 } }$ \\
			\hline
			\multirow{5}{*}{-0.5} & 2 pairs & ${ { 1.4789 \text{e} }^{ -5 } }$ & ${ { 1.5996 \text{e} }^{ -5 } }$ & ${ { 1.6549 \text{e} }^{ -5 } }$ & ${ { 1.6985 \text{e} }^{ -5 } }$ & ${ { 1.7339 \text{e} }^{ -5 } }$ \\
			& 3 pairs & ${ { 4.1116 \text{e} }^{ -5 } }$ & ${ { 4.4524 \text{e} }^{ -5 } }$ & ${ { 4.6087 \text{e} }^{ -5 } }$ & ${ { 4.7321 \text{e} }^{ -5 } }$ & ${ { 4.8322 \text{e} }^{ -5 } }$ \\
			& 4 pairs & ${ { 9.6271 \text{e} }^{ -5 } }$ & ${ { 1.0434 \text{e} }^{ -4 } }$ & ${ { 1.0810 \text{e} }^{ -4 } }$ & ${ { 1.1109 \text{e} }^{ -4 } }$ & ${ { 1.1351 \text{e} }^{ -4 } }$ \\
			& 5 pairs & ${ { 2.5972 \text{e} }^{ -4 } }$ & ${ { 2.8007 \text{e} }^{ -4 } }$ & ${ { 2.9096 \text{e} }^{ -4 } }$ & ${ { 2.9961 \text{e} }^{ -4 } }$ & ${ { 3.0670 \text{e} }^{ -4 } }$ \\
			\hline
			\multirow{5}{*}{-0.7} & 2 pairs & ${ { 6.8983 \text{e} }^{ -5 } }$ & ${ { 7.4729 \text{e} }^{ -5 } }$ & ${ { 7.7371 \text{e} }^{ -5 } }$ & ${ { 7.9457 \text{e} }^{ -5 } }$ & ${ { 8.1148 \text{e} }^{ -5 } }$ \\
			& 3 pairs & ${ { 1.9074 \text{e} }^{ -4 } }$ & ${ { 2.0725 \text{e} }^{ -4 } }$ & ${ { 2.1486 \text{e} }^{ -4 } }$ & ${ { 2.2089 \text{e} }^{ -4 } }$ & ${ { 2.2579 \text{e} }^{ -4 } }$ \\
			& 4 pairs & ${ { 4.3862 \text{e} }^{ -4 } }$ & ${ { 4.7885 \text{e} }^{ -4 } }$ & ${ { 4.9771 \text{e} }^{ -4 } }$ & ${ { 5.1275 \text{e} }^{ -4 } }$ & ${ { 5.2507 \text{e} }^{ -4 } }$ \\
			& 5 pairs & ${ { 1.1595 \text{e} }^{ -3 } }$ & ${ { 1.2756 \text{e} }^{ -3 } }$ & ${ { 1.3361 \text{e} }^{ -3 } }$ & ${ { 1.3849 \text{e} }^{ -3 } }$ & ${ { 1.4254 \text{e} }^{ -3 } }$ \\
		\end{tabular}
	\end{ruledtabular}
\caption{\label{table_1}Comparison between exact GSM diagonalization and Richardson calculation using Eqs. (\ref{richardson_equations}). The relative error of the total energy (\ref{total_energy}) calculated using Eqs. (\ref{richardson_equations}) is shown for various pairing strengths ${ G }$, different number of fermion pairs and different number of discretization points along the real-energy contour. }
\end{table*}	
		Each level is doubly degenerate, {\em i.e.} there is only one pair of fermions per level. To assure the completeness of the s.p. basis, the set of s.p. states from the discretized real-energy contour is added.
		The contour is composed of three segments: $[k_0 ; k_1]=[0.0 ; 0.5]$, $[k_1 ; k_2]=[0.5 ; 1.0]$, and $[k_2 ; k_{\rm max}]=[1.0 ; 2.0]$, and the calculations are performed for different strengths ${ G }$ of the pairing interaction: ${ G = 0.01 }$ MeV, ${ G = 0.3 }$ MeV, ${ G = 0.5 }$ MeV and ${ G = 0.7 }$ MeV.		      The Gauss-Legendre method is used to select optimal discretized s.p. levels along the real-energy contour for each given number of the discretization points.
		The same set of s.p. levels and the corresponding Gaussian weights are then used to find the total energy of the system by solving both, the generalized Richardson equation (\ref{richardson_equations}) and the GSM.
		The relative error of the total energy ${ \mathcal{E} }$ (the real part of ${ \tilde{ \mathcal{E}} }$ in Eq. (\ref{total_energy}))
		calculated using Eqs. (\ref{richardson_equations}) with respect to the exact GSM energy: ${ \delta ( \mathcal{E} ) = ({ \mathcal{E} }_{ \text{GSM} } - \mathcal{E} ) / { \mathcal{E} }_{ \text{GSM} } }$, is shown in Table \ref{table_1} for different total number of the discretization points.
		Each segment of the contour ${ { L }_c^{ + } }$ is discretized with the same number of points. One may notice that the discrepancy between GSM and generalized Richardson results grows with increasing the pairing strength and number of fermion pairs.
		The expression (\ref{richardson_equations}) for the pair energies does not account accurately for the pair-pair interaction due to the approximation made in the commutators (\ref{commutator_discrete_continuum}). As expected energy obtained by solving generalized Richardson equations for a single pair (\ref{richardson_equations}) coincide with the exact GSM result.
		}

	\subsubsection{Weakly bound and resonances states}
		{

		The evolution of the relative error of the generalized Richardson equations (\ref{richardson_equations}) for weakly bound and resonance double degenerate s.p. levels is shown in this subsection as a function of the pairing strength for 2 and 3 pairs of fermions. Different spectra of s.p. pole states used in these calculations are shown in Table \ref{table_spectra}.
		\begin{table}[h!]
		\begin{ruledtabular}
		\begin{tabular}{cc}
			Spectrum  & Single-particle energies (MeV) \\
			\hline \\
			1 & \{ -2.5 , -1.5 , -0.5 \} \\
			2 & \{ -1.5 , -0.5 , (0.5 , -0.05) \} \\
		        3 & \{ -0.5 , (0.5 , -0.05) , (1.5 , -0.15) \} \\
			4 & \{ -2.5 , -1.5 , -0.5 , (0.5 , -0.05) \} \\
		\end{tabular}
		\end{ruledtabular}
		\caption{\label{table_spectra} The s.p. levels used in the studies of the relative error of the generalized Richardson approach (\ref{richardson_equations}).}
		\end{table}		
		To construct the complete Berggren s.p. basis, we take for each considered resonance state a different contour in the complex $k$-plane.
		The contour used for the spectrum 1 in Table \ref{table_spectra} is divided into three segments along the real-$k$ axis: $[k_0 ; k_1]=[0.0 ; 0.5]$, $[k_1 ; k_2]=[0.5 ; 1.0]$, and $[k_2 ; k_{\rm max}]=[1.0 ; 2.0]$. The parametrization of contours for different resonances is shown in Table \ref{table_contours}.	Each contour is discretized with 30 points selected by the Gauss-Legendre quadrature procedure and all segments are discretized with 10 points.	
		
		\begin{table}[h!]
		\begin{ruledtabular}
		\begin{tabular}{ccccc}
			Resonance  & $k_0 \; ({ \text{fm} }^{ -1 }) $ & ${ k }_1 \; ({ \text{fm} }^{ -1 }) $ & ${ k }_2 \; ({ \text{fm} }^{ -1 }) $ & ${ k }_{ \text{max} } \; ({ \text{fm} }^{ -1 }) $ \\
			\hline \\
			${ (0.5 , -0.05) }$ & $0.0$ & (0.1549 , -0.14) & 1.0 & 2.0 \\
			${ (1.5 , -0.15) }$ & $0.0$ & (0.2682 , -0.2) & 1.0 & 2.0 \\
		\end{tabular}
		\end{ruledtabular}
		\caption{\label{table_contours}Parameters of the contours in the complex-$k$ plane associated with the resonance poles.}
		\end{table}	
			
			\begin{figure}[]
		\includegraphics[angle=00,width=0.8\columnwidth]{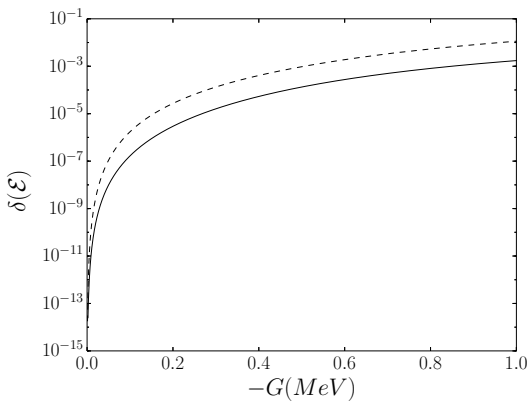}
		\caption{Spectrum 1 (Table \ref{table_spectra}): The relative error ${ \delta (\mathcal{E}) }$ of the ground state energy (\ref{total_energy}) calculated using the pair energies ${ { E }_{ i } }$ given by the generalized Richardson equations  (\ref{richardson_equations}) is plotted as a function of the pairing strength $G$. Results for two (three) pairs of fermions are shown with the red (blue) line.}
		\label{graph_spectrum_1}
		\end{figure}

		\begin{figure}[]
			\vskip 1truecm
\includegraphics[angle=00,width=0.8\columnwidth]{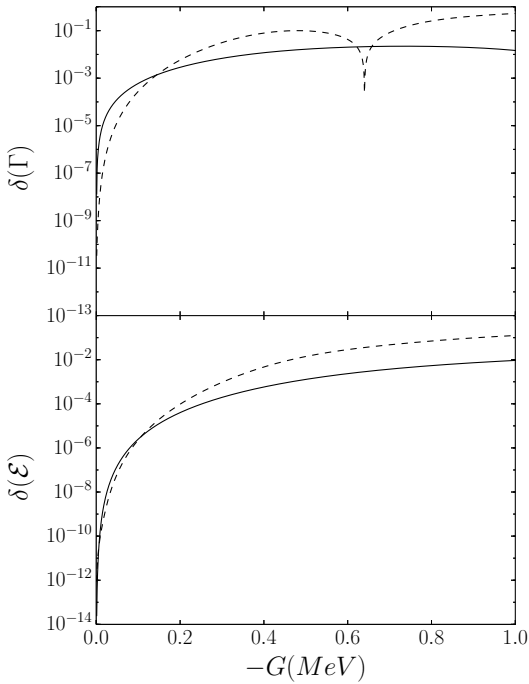}
\caption{Spectrum 2 (Table \ref{table_spectra}): The relative error of the ground state energy ${ \delta (\mathcal{E}) }$ and width ${ \delta (\Gamma) }$ which are calculated using the pair energies ${ { E }_{ i } }$ given by the generalized Richardson equations (\ref{richardson_equations}). For more details, see the caption of Fig. \ref{graph_spectrum_1}. }
		\label{graph_spectrum_2}
		\end{figure}

		\begin{figure}[]
			\vskip 1truecm
		\includegraphics[angle=00,width=0.8\columnwidth]{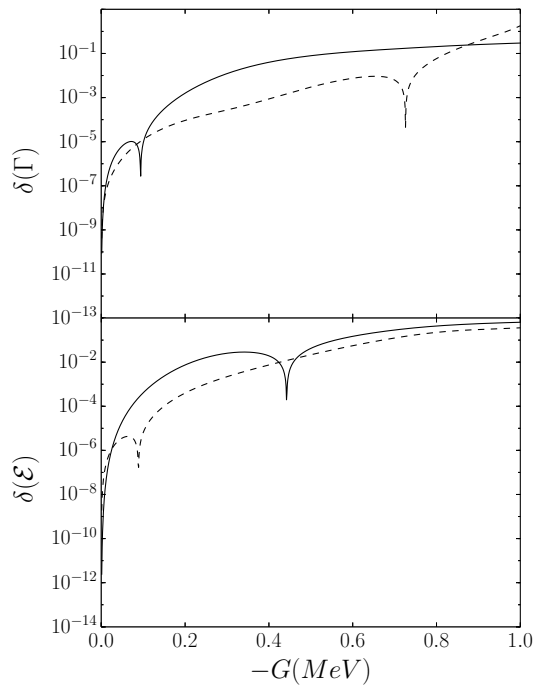}
		\caption{The same as in Fig. \ref{graph_spectrum_2} but for spectrum 3 in Table \ref{table_spectra}. }
		\label{graph_spectrum_3}
		\end{figure}

		\begin{figure}[]
			\vskip 1truecm
		\includegraphics[angle=00,width=0.8\columnwidth]{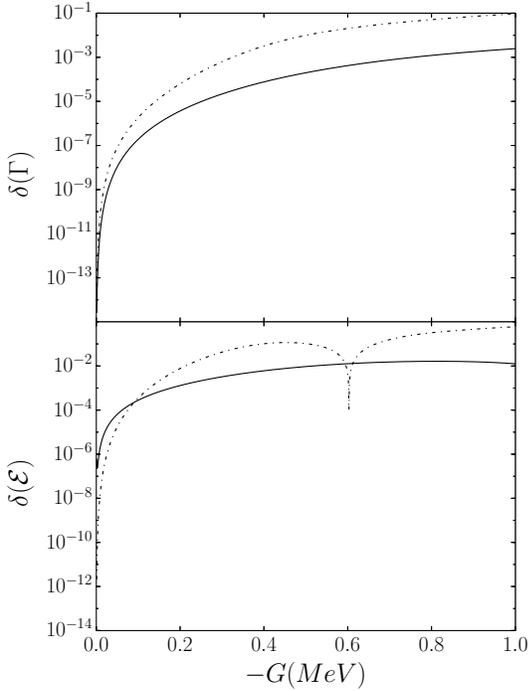}
		\caption{The same as in Fig. \ref{graph_spectrum_2} but for spectrum 4 in Table \ref{table_spectra}. The contour in the complex-$k$ plane for the resonance pole at $(0.5 {\rm MeV}, -0.05 {\rm MeV})$ is: $[k_0 ; k_1]=[0.0 ; (0.1549, -0.2)]$, $[k_1 ; k_2]=[(0.1549, -0.2) ; 1.0]$, and
	$[k_2 ; k_{\rm max}]=[1.0 ; 2.0]$.  Results for two (four) pairs of fermions are shown with red (blue) line.
		}
		\label{graph_spectrum_4}
		\end{figure}

		The dependence on the pairing strength $G$ of the relative error of the ground state energy and the width calculated using the generalized Richardson approach is plotted in Figs. \ref{graph_spectrum_1} to \ref{graph_spectrum_4} for different s.p. spectra shown in Table \ref{table_spectra}. Numerical results show a strong dependence of the relative error on the pairing strength and the number of fermion pairs. One may also notice (see Figs. \ref{graph_spectrum_2} - \ref{graph_spectrum_4}) few spikes of the relative error for the ground state energy and/or the width at certain values of the pairing strength. At these discrete values of $G$, either real or imaginary part of the complex total energy (\ref{total_energy}) calculated using the generalized Richardson approach (\ref{richardson_equations}) is equal to the GSM energy. We found these spikes in ${ \delta (\mathcal{E}) }$ and/or ${ \delta (\Gamma) }$ only in the cases of s.p. spectra with at least one resonance.
		
		In Table \ref{spectrum_study} we present the relative error of the total energy of all discrete states of the Hamiltonian (\ref{new_pairing_hamiltonian}) for two values of the pairing strength: ${ G = -0.4 }$ MeV and ${ G = -0.7 }$ MeV. We take three pairs of fermions and the s.p. spectrum is given by five doubly degenerate levels with energies: ${ { \epsilon }_{ i } = \{ -2.5 , -1.5 , -0.5 , (0.5,-0.05) , (1.5,-0.15) \} }$ in units of MeV. The s.p. contours in the $k$-plane are given in Table \ref{table_contours}. In this case, there are ten different discrete many-body pole states.
		\begin{table}[!h]
		\begin{ruledtabular}
		\vskip 1 truecm
		\begin{tabular}{cc|cc|cc}
		&   & \multicolumn{2}{c}{ ${ G = -0.4 }$MeV } & \multicolumn{2}{c}{ ${ G = -0.7 }$MeV } \\
		State & Conf & ${ \delta (\mathcal{E}) }$ & ${ \delta (\Gamma) }$ & ${ \delta (\mathcal{E}) }$ & ${ \delta (\Gamma) }$ \\
			\hline
			1  & 11100 & ${ { 8.2900 \text{e} }^{ -4 } }$ & ${ { 1.6856 \text{e} }^{ -2 } }$ &  ${ { 6.3687 \text{e} }^{ -3 } }$ & ${ { 2.0974 \text{e} }^{ -2 } }$  \\
			2  & 11010 & ${ { 5.8954 \text{e} }^{ -4 } }$ & ${ { 4.9297 \text{e} }^{ -2 } }$ &  ${ { 3.9907 \text{e} }^{ -3 } }$ & ${ { 2.9677 \text{e} }^{ -1 } }$  \\
			3  & 11001 & ${ { 7.6322 \text{e} }^{ -5 } }$ & ${ { 1.5083 \text{e} }^{ -3 } }$ &  ${ { 7.0742 \text{e} }^{ -4 } }$ & ${ { 1.3266 \text{e} }^{ -2 } }$  \\
			4  & 10110 & ${ { 2.5319 \text{e} }^{ -3 } }$ & ${ { 5.1784 \text{e} }^{ -2 } }$ &  ${ { 2.3193 \text{e} }^{ -2 } }$ & ${ { 1.3863 \text{e} }^{ -1 } }$  \\
			5  & 01110 & ${ { 6.2516 \text{e} }^{ -3 } }$ & ${ { 6.1601 \text{e} }^{ -2 } }$ &  ${ { 4.3516 \text{e} }^{ -2 } }$ & ${ { 1.7037 \text{e} }^{ -2 } }$  \\
			6  & 10101 & ${ { 1.5258 \text{e} }^{ -4 } }$ & ${ { 1.3426 \text{e} }^{ -3 } }$ &  ${ { 1.7037 \text{e} }^{ -2 } }$ & ${ { 1.2335 \text{e} }^{ -1 } }$  \\
			7  & 10011 & ${ { 2.2406 \text{e} }^{ -4 } }$ & ${ { 1.7029 \text{e} }^{ -4 } }$ &  ${ { 8.7716 \text{e} }^{ -4 } }$ & ${ { 5.7504 \text{e} }^{ -3 } }$  \\
			8  & 01101 &  ${ { 2.2482 \text{e} }^{ -4 } }$ & ${ { 1.5166 \text{e} }^{ -3 } }$ & ${ { 1.1971 \text{e} }^{ -4 } }$ & ${ { 6.1501 \text{e} }^{ -3 } }$  \\
			9  & 01011 & ${ { 2.4802 \text{e} }^{ -2 } }$ & ${ { 1.2426 \text{e} }^{ -3 } }$ &  ${ { 1.3286 \text{e} }^{ -2 } }$ & ${ { 7.7301 \text{e} }^{ -3 } }$  \\
			10  & 00111 & ${ { 6.9734 \text{e} }^{ -4 } }$ & ${ { 7.9498 \text{e} }^{ -4 } }$ & ${ { 1.6944 \text{e} }^{ -2 } }$ & ${ { 5.6188 \text{e} }^{ -3 } }$  \\
		\end{tabular}
		\end{ruledtabular}
		\caption{\label{spectrum_study}The relative error of the complex energy (\ref{total_energy}) for all excited states of a pairing Hamiltonian (\ref{new_pairing_hamiltonian}) with three pairs of fermions distributed over five doubly degenerate levels and three discretized continua. The pole space configuration for each state, {\em i.e.} the occupation by pairs of fermions of each discrete s.p. level, is indicated in the second column for $G=0$. For more details, see the description in the text.}
		\end{table}		
As one can see in Table \ref{spectrum_study}, precision of the calculation using the generalized Richardson approach (\ref{richardson_equations}) can vary by two orders of magnitude from one state to another and no simple tendency with increasing the excitation energy can be noticed. For that reason, also the relative error of the transition energy between neighboring states varies from one state to another in the unpredictable way.  As a rule, the relative error for the imaginary part of the total energy is bigger than the corresponding error of the real part.

		\begin{figure}[]
		\vskip 1truecm
		\includegraphics[angle=00,width=0.7\columnwidth]{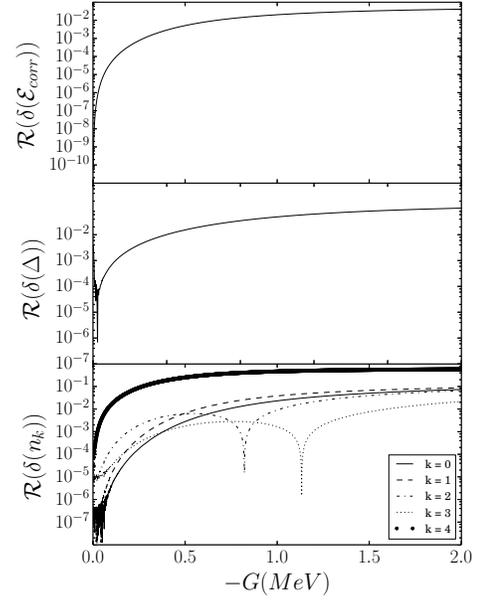}
		\caption{Spectrum 2 (Table \ref{table_spectra}): The relative error of the generalized Richardson solution for real parts of: (i) the correlation energy ${ { \mathcal{E} }_{ corr } }$ (the upper part), (ii) the pairing gap ${ \Delta }$ (the middle part), and (iii) the occupation probability ${ { n }_{ k } }$ for 5 lowest s.p. states $K=0\dots,4$ (the lower part). These calculations have been performed for the ground state of the spectrum 2. }
		\label{compil_gs}
		\end{figure}
		
		\begin{figure}[]
		\vskip 1truecm
		\includegraphics[angle=00,width=0.7\columnwidth]{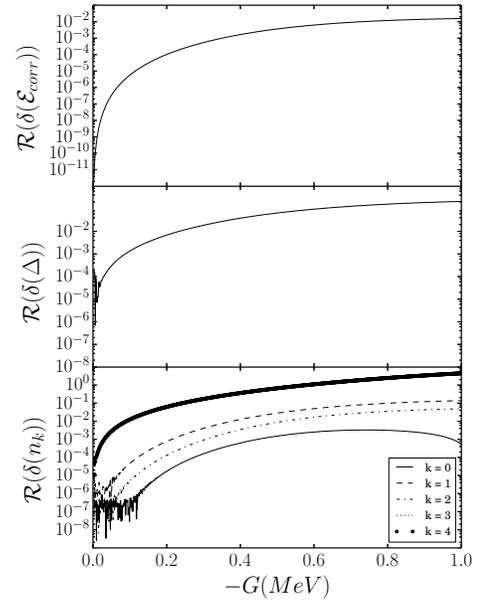}
		\caption{The same as in Fig. \ref{compil_gs} but for the first excited state.}
		\label{compil_1st}
		\end{figure}

		\begin{figure}[]
		\vskip 1truecm
		\includegraphics[angle=00,width=0.7\columnwidth]{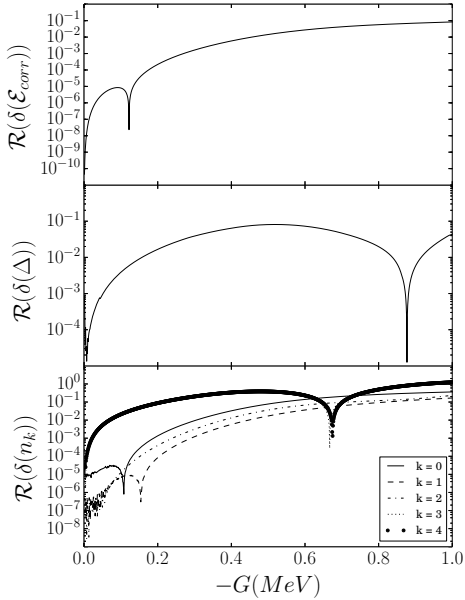}
		\caption{The same as in Fig. \ref{compil_gs} but for the second excited state.}
		\label{compil_2nd}
		\end{figure}		
	
		\begin{figure}[]
		\vskip 1truecm
		\includegraphics[angle=00,width=0.7\columnwidth]{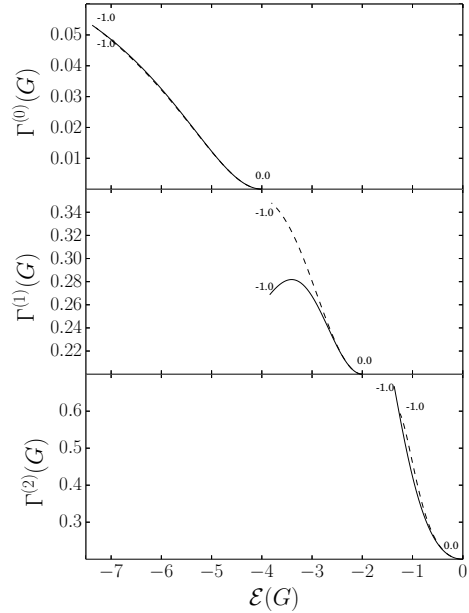}
		\caption{Spectrum 2 (Table \ref{table_spectra}): The evolution of the three lowest complex eigenvalues
		$(({\mathcal{E}^{(i)},\Gamma^{(i)}})~;~ i=0,1,2$) of the pairing Hamiltonian is plotted as function of the pairing strength 
		for 2 pairs of fermions. The upper most (lowest) figure shows results for the second excited (ground) state, whereas the 
		figure in the middle is for the first excited state. The solid and dashed lines show the exact GSM solution, and the solution
		 (\ref{total_energy}) of the generalized Richardson approach (\ref{richardson_equations}), respectively. Numbers at the 
		 curves denote limiting values of the pairing strength (in MeV).}
		\label{complex_energy_fig}
		\end{figure}		

		\begin{figure}[]
		\vskip 1truecm
		\includegraphics[angle=00,width=0.7\columnwidth]{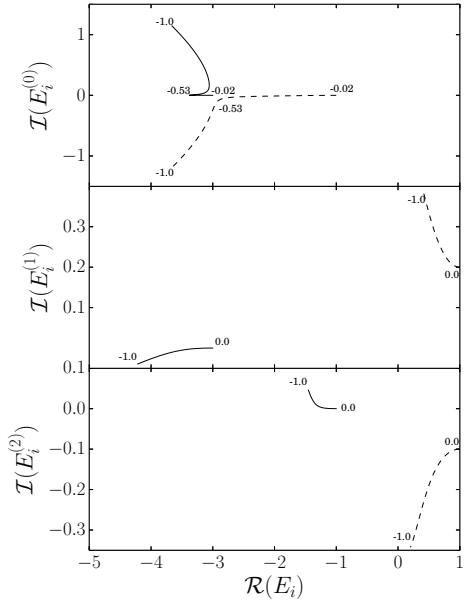}
		\caption{Spectrum 2 (Table \ref{table_spectra}): The evolution of the lowest two (complex) pair energies ${ { E }_{ i }^{ (K) } }$ (${ i = 0,1 }$) with the pairing strength for 2 pairs of fermions. The pair energies are obtained by solving the generalized Richardson equations (\ref{richardson_equations}) for the ground state ${ (K = 0) }$, and for the two lowest excited states (${ K = 1,2 }$). Numbers at the curves show limiting values of the pairing strength (in MeV). }
		\label{pair_energies_fig}
		\end{figure}		

		In Figs. \ref{compil_gs}, \ref{compil_1st}, and \ref{compil_2nd}, we present the relative error for other relevant quantities: the correlation energy ${ { \mathcal{E} }_{ corr } }$, the pairing gap ${ \Delta }$, and the occupation probability ${ { n }_{ k } }$ for 5 lowest s.p. states $(k=0,\dots,4)$. The calculations are performed for two pairs of fermions. Results are shown for the ground state, and the next two excited states. The correlation energy is calculated as: ${ { \mathcal{E} }_{ corr } = { \mathcal{E} }_{ G = 0 } - \mathcal{E} }$. The pairing gap ${ \Delta }$ is calculated according to Eq. (\ref{gap_expression}). In GSM, the occupation probabilities are determined using Eq. (\ref{nk_exact}), whereas in the generalized Richardson equations approach we use Eq. (\ref{nk_richardson}).  One can see that deeps in the relative error of different quantities shown in Figs. \ref{compil_gs}-\ref{compil_2nd}, do not appear at the same values of the pairing strength.
	
		The trajectory of complex eigenvalues (${ \mathcal{E}-\Gamma }$) of the pairing Hamiltonian in the energy-width plane is plotted in Fig. \ref{complex_energy_fig} as a function of the pairing strength ${ G }$ in the interval from 0 to -1 MeV for the ground state $(K=0)$ (the upper part), the first excited state $(K=1)$ (the middle part), and the second excited state ($K=2$) (the lower part) excited state. The solid (dashed) lines show the solutions of GSM (generalized Richardson equations). One may notice that the relative discrepancy between exact and approximate results is largest for the first excited state at large values of the pairing strength $G$ .
		
		In Fig. \ref{pair_energies_fig}, the trajectory of pair energies in the complex energy plane is plotted for the ground state $(K=0)$ (the upper part), the first excited state $(K=1)$ (the middle part), and the second excited state $(K=2)$  as a function of the pairing strength ${ G }$ in the interval from 0 to -1 MeV. In the upper part of the figure, one can see that the pair energies in an interval $0>G>-0.53$ MeV tend to approach each other along the real-energy axis. At $G \sim -0.53$ MeV, these two pair energies exhibit an avoided crossing and then move rapidly into the complex-energy plane with increasing value of the pairing strength. The pattern of avoided crossings, {\em i.e.} mixing pair energies, is a general pattern and can be seen for excited states $(K=1,2)$ as well.
		
		}
}

	\subsection{Application of generalized Richardson equations to physical systems}
	\label{applications_physical_system}
	{
	
	In the previous sections, we solved the generalized Richardson equation for the rational Gaudin model with the continuum. In order to obtain the Richardson-like solution for this generalized pairing problem, we had to compromise commutation relations for the non-resonant continuum states.  Therefore, whenever the occupation of non-resonant continuum states becomes important, one might expect that the solution of the generalized Richardson equation is less accurate. This happens for strong pairing correlations.
	
	To test this expectation, we compared solutions of the generalized Richardson equation with exact GSM solutions.
		We have shown that even though the relative error of the generalized Richardson solution growth with the number of fermion pairs and the pairing strength, nevertheless it remains rather accurate, especially in the limit of weak pairing correlations. One can use this model to simulate various situations involving pairing correlations and continuum in weakly bound or unbound states. In particular, one can use this model to test the common strategy of nuclear SM to replace effects of continuum couplings by the phenomenological adjustment of both s.p. energies and two-body matrix elements.
		
		Like many well-known group theoretical models developed in nuclear physics, the rational Gaudin model with the continuum can be applied to calculate not only energy spectra but also transitions probabilities in the long series of isotopes. One should stress however that the absence of particle-hole interaction makes this model unrealistic, as the essential element of the competition between pairing and quadrupole interaction is missing.
			
Below, we will apply generalized Richardson equations to calculate spectra of carbon isotopes and investigate the role of the continuum in these spectra. We will also comment on a possibility to investigate the weak-pairing limit of the ultra-small superconducting grains which is characterized by strong fluctuations of the pairing field.

	\subsubsection{Spectra of carbon isotopes}
	\label{applications_carbon_isotopes}
	{
		
		To illustrate possible applications of the generalized Richardson equations, we will now calculate spectra of carbon isotopes with $14 \leq A \leq 20$. The choice of parameters in the Hamiltonian (\ref{new_hamiltonian_discretized}) is motivated by the experimental spectrum of ${ {  }^{ 13 } }$C and the binding energy of ${ {  }^{ 14 } }$C. In this calculation, we assume the core of ${ {  }^{ 12 } }$C and calculate energies of all states in $^{14-20}$C with respect to the energy of this core.
	
	Berggren basis consists of the pole s.p. states: ${ 0{ p }_{ 1/2 } }$, ${ 1{ s }_{ 1/2 } }$, ${ 0{ d }_{ 5/2 } }$, ${ 0 { d }_{ 3/2 } }$, 
	${ 0{ f }_{ 7/2 } }$, and the two non-resonant continua: $\{ { d }_{ 3/2 } \}$, $\{ { f }_{ 7/2 } \}$. S.p. energies of bound states 
	${ 0{ p }_{ 1/2 } }$, ${ 1{ s }_{ 1/2 } }$, ${ 0{ d }_{ 5/2 } }$ are given by experimental energies of 
	${ { 1/2 }^{ - }_1, { 1/2 }^{ + }_1 }$ and ${ { 5/2 }^{ + }_1 }$ states in ${ {  }^{ 13 } }$C: 
	${ { \epsilon }_{ 0{ p }_{ 1/2 } } = -4.946 }$ MeV, ${ { \epsilon }_{ 1{ s }_{ 1/2 } } = -1.857 }$ MeV, and 
	${ { \epsilon }_{ 0{ d }_{ 5/2 } } = -1.093 }$ MeV. The energy of resonances ${ 0 { d }_{ 3/2 } }$ and ${ 0 { f }_{ 7/2 } }$ 
	are \cite{idbetan12_7}: ${ { \epsilon }_{ 0{ d }_{ 3/2 } } = (2.267~{\rm MeV};-0.416~ {\rm MeV}) }$ and \\
	${ { \epsilon }_{ 0{ f }_{ 7/2 } } = (9.288~{\rm MeV};-3.040~{\rm MeV}) }$.
	The complex contours $\{ { d }_{ 3/2 } \}$ and $\{ { f }_{ 7/2 } \}$ associated with ${ 0 { d }_{ 3/2 } }$ and 
	${ 0 { f }_{ 7/2 } }$ resonance are given in Table \ref{contour_carbon}. They are discretized with 10 points per segment, 
	{\em i.e.} 30 points per contour.
	\begin{table}[h!]
	\centering
	\begin{tabular}{ccccc}
		Resonance  & $k_0 \; ({ \text{fm} }^{ -1 }) $ & ${ k }_1 \; ({ \text{fm} }^{ -1 }) $ & ${ k }_2 \; ({ \text{fm} }^{ -1 }) $ & ${ k }_{ \text{max} } \; ({ \text{fm} }^{ -1 }) $ \\
		\hline \\
		${ { d }_{ 3/2 } }$ & $0.0$ & (0.332 , -0.03) & 0.66 & 2.0 \\
		${ { f }_{ 7/2 } }$ & $0.0$ & (0.678 , -0.1) & 1.24 & 2.0 \\
	\end{tabular}
	\caption{Parameters of the contours $L^+$ in the complex $k$-plane, associated with ${ 0 { d }_{ 3/2 } }$ and ${ 0 { f }_{ 7/2 } }$ resonance poles. Each contour consists of three segments: $\left[ k_0,k_1 \right]$, $\left[ k_1,k_2 \right]$, $\left[ k_2,k_{\rm max} \right]$, and each segment is discretized with 10 points.
	}
	\label{contour_carbon}
	\end{table}	
For the pairing strength, we take: ${ G = \chi / A }$, where $\chi=-11.13$ MeV. The constant $\chi$ is adjusted to reproduce the experimental binding energy of ${ {  }^{ 14 } }$C with respect to ${ {  }^{ 12 } }$C.

To evaluate the role of the continuum in the spectra of carbon isotopes, we compare results of the generalized Richardson equations (\ref{richardson_equations}) with results of the standard Richardson calculations (\ref{richardson_equations}) without continuum couplings and with real s.p. energies. In the latter case, the s.p. energies of the bound states: ${ 0{ p }_{ 1/2 } }$, ${ 1{ s }_{ 1/2 } }$, ${ 0{ d }_{ 5/2 } }$, are the same as given above, and energies of ${ {0d }_{ 3/2 } }$ and ${ { 0f }_{ 7/2 } }$ resonances are real: ${ { \epsilon }_{ 0{ d }_{ 3/2 } } = 2.267~{\rm MeV} }$ and ${ { \epsilon }_{ 0{ f }_{ 7/2 } } = 9.288~{\rm MeV} }$. To reproduce the experimental binding energy of ${ {  }^{ 14 } }$C in this SM-like basis, the pairing strength is increased $\chi=-15.064$ MeV.

	In Table \ref{exp_gs}, we compare experimental binding energies (${ { B }_{\rm exp } }$) with binding energies calculated
	using either generalized Richardson equations
	($B_{\rm GR}$) or standard Richardson equations which neglect continuum effects ($B_{\rm R}$). All energies
	are given with respect to the energy of ${ {  }^{ 12 } }$C.
        \begin{table}[h!]
	\centering
	\begin{tabular}{cccc}
		Isotope & ${ { B }_{\rm exp } }$ (MeV) & ${ { B }_{ \rm GR } }$ (MeV) &${ { B }_{\rm R } }$ (MeV)\\
		\hline \\
		${ {  }^{ 14 } }$C & 13.123  & 13.124 & 13.124 \\
		${ {  }^{ 16 } }$C & 18.590 & 20.814 & 20.477 \\
		${ {  }^{ 18 } }$C & 23.505 & 25.130 & 24.386 \\
		${ {  }^{ 20 } }$C & 27.013 & 27.170 & 25.886 \\
	\end{tabular}
	\caption{Binding energy in the chain of carbon isotopes $^{14-20}$C. ${ { B }_{\rm  GR } }$ and  ${ { B }_{\rm  R } }$ give results of the generalized Richardson equations (\ref{richardson_equations}) and standard Richardson equations (\ref{richardson_equations}), respectively. ${ { B }_{\rm exp } }$ gives the experimental binding energy. All energies are given with respect to the energy of ${ {  }^{ 12 } }$C.
	}
	\label{exp_gs}
	\end{table}	
	One can see that continuum changes the $A$-dependence of  binding energies.
		Interestingly, $B_{\rm GR}$ is equal to ${ { B }_{\rm exp } }$ both in $^{14}$C and in $^{20}$C.
	\begin{figure}[ht]
		\centering
		\includegraphics[width=0.7\linewidth]{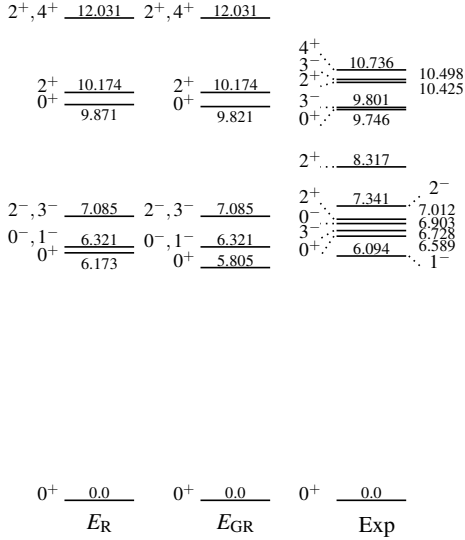}
		\caption{The experimental spectrum of ${ {  }^{ 14 } }$C is compared with the spectra calculated using either the standard Richardson equations (no continuum) $(E_{\rm R}$), or and generalized Richardson equations $(E_{\rm GR}$). For more details, see the discussion in the text.}
	\label{fig_14C_spectrum}
	\end{figure}
\begin{table}[h!]
	\centering
	\begin{tabular}{cccc}
		Conf & State & ${ { E }_{ \text{GR} } } ({\rm MeV})$ & ${ { E }_{ \text{R} } } ({\rm MeV})$  \\
		\hline \\
		${ { (1) }^{ 2 } }$ & ${ { 0 }^{ + } }$ & 0  & 0  \\
		${ { (2) }^{ 2 } }$ & ${ { 0 }^{ + } }$ & 5.805  & 6.173  \\
		${ (1)^1(2)^1 }$ & ${ { 0 }^{ - }, { 1 }^{ - } }$ & 6.321  & 6.321  \\
		${ (1)^1(3)^1 }$ & ${ { 2 }^{ - }, { 3 }^{ - } }$ & 7.085  & 7.085  \\
		${ { (3) }^{ 2 } }$ & ${ { 0 }^{ + } }$ & 9.821  & 9.871  \\
		${ (2)^1(3)^1 }$ & ${ { 2 }^{ + } }$ & 10.174  & 10.174  \\
		${ (3)^1(3)^1 }$ & ${ { 2 }^{ + }, { 4 }^{ + } }$ & 12.031  & 12.031  \\
	\end{tabular}
	\caption{The initial configuration ($G=0$) and excitation energies of different states of ${ {  }^{ 14 } }$C calculated using both the generalized Richardson equations ($E_{\rm GR}$) and the standard Richardson ($E_{\rm R}$) equations. The initial configuration is denoted by the index of an occupied level $(1\equiv 0p_{1/2}, 2\equiv 1s_{1/2}, 3\equiv 0d_{5/2})$ and the number of particles in a given level (1 or 2). }
\label{carbon14_conf}
	\end{table}
	\begin{figure}[ht]
		\centering
		\includegraphics[width=0.7\linewidth]{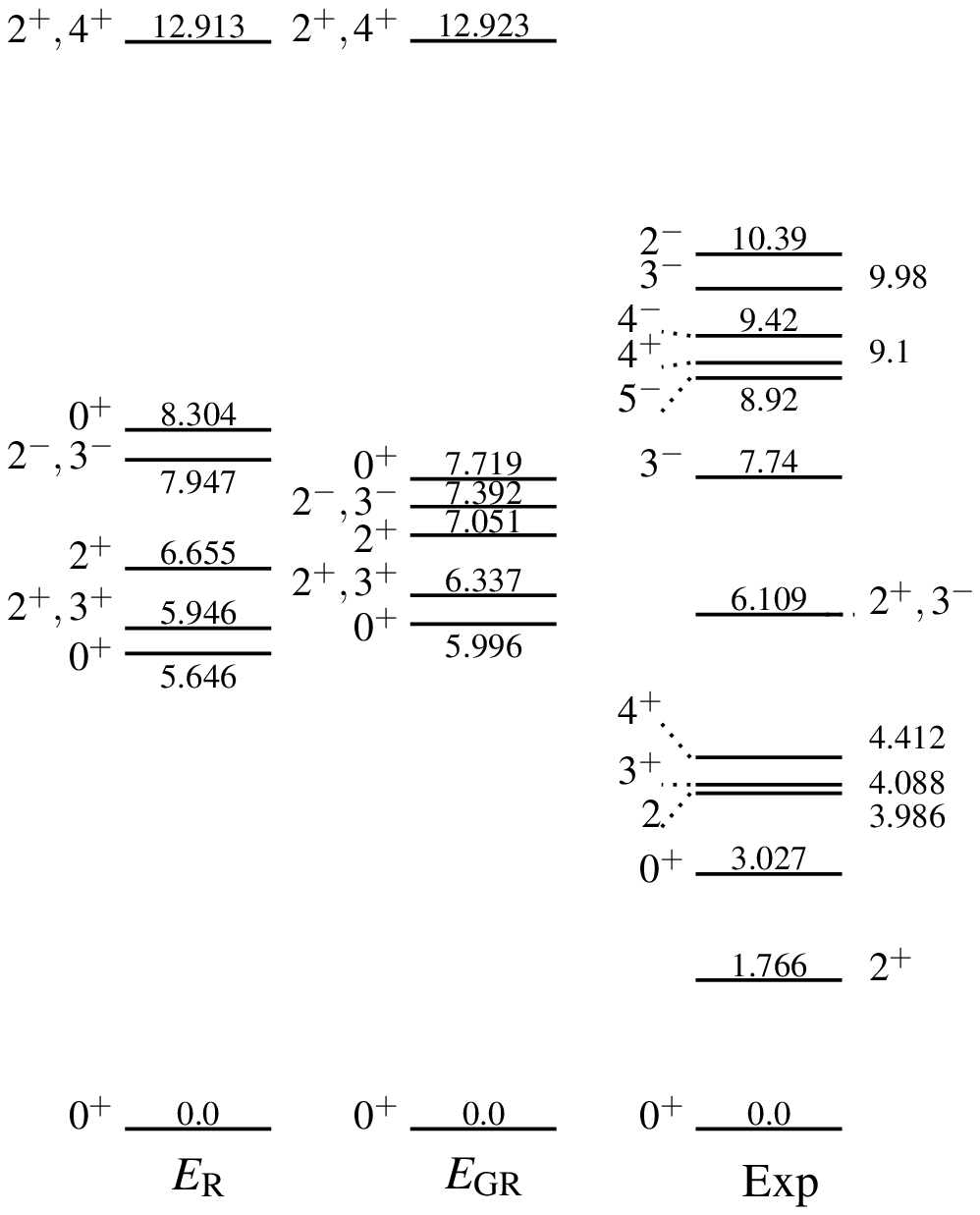}
		\caption{Experimental spectrum of ${ {  }^{ 16 } }$C is compared with the spectra calculated using either the standard Richardson equations (no continuum) $(E_{\rm R}$), or and generalized Richardson equations $(E_{\rm GR}$). For more details, see the discussion in the text.		}
	\label{fig_16C_spectrum}
	\end{figure}
\begin{table}[h!]
	\centering
	\begin{tabular}{ccccc}
	Conf & State & ${ { E }_{ \text{GR} } } ({\rm MeV})$ & ${ { E }_{ \text{R} } } ({\rm MeV})$  \\
		\hline \\
		${ { (1) }^{ 2 } { (2) }^{ 2 } }$ & ${ { 0 }^{ + } }$ & 0 & 0  \\
		${ { (1) }^{ 2 } { (3) }^{ 2 } }$ & ${ { 0 }^{ + } }$ & 5.996  & 5.646  \\
		${ { (1) }^{ 2 } (2)^1(3)^1 }$ & ${ { 2 }^{ + }, { 3 }^{ + } }$ & 6.337  & 5.946  \\
		${ { (1) }^{ 2 } (3)^1(3)^1 }$ & ${ { 2 }^{ + }, { 4 }^{ + } }$ & 7.051  & 6.655  \\
		${ { (2) }^{ 2 } (1)^1(3)^1 }$ & ${ { 2 }^{ - }, { 3 }^{ - } }$ & 7.392  & 7.947  \\
		${ { (2) }^{ 2 } { (3) }^{ 2 } }$ & ${ { 0 }^{ + } }$ & 7.719  & 8.304  \\
		${ { (2) }^{ 2 } (3)^1(3)^1 }$ & ${ { 2 }^{ + }, { 4 }^{ + } }$ & 12.923  & 12.913  \\
	\end{tabular}
	\caption{The initial configuration ($G=0$) and energies of different states of ${ {  }^{ 16 } }$C calculated using both the generalized Richardson equations ($E_{\rm GR}$) and the standard Richardson ($E_{\rm R}$). For details, see the caption of Fig. \ref{carbon14_conf}. }
	\label{carbon16_conf}
	\end{table}		
	\begin{figure}[ht]
		\centering
		\includegraphics[width=0.5\linewidth]{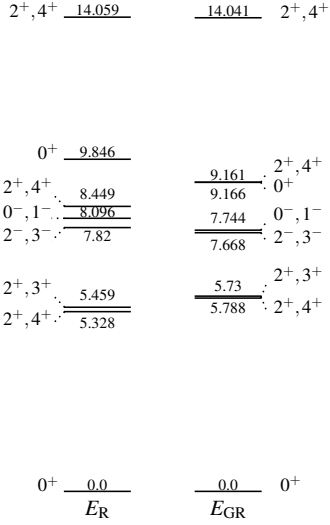}
		\caption{The spectrum of ${ {  }^{ 18 } }$C calculated using either the standard Richardson equations (no continuum) $(E_{\rm R}$), or the generalized Richardson equations $(E_{\rm GR}$). For more details, see the discussion in the text. We omitted the second ${ { 0 }^{ + } }$ as mentioned in Tab. \ref{carbon18_conf}}
	\label{fig_18C_spectrum}
	\end{figure}
	\begin{table}[h!]
	\centering
	\begin{tabular}{ccccc}
	Conf & State & ${ { E }_{ \text{GR} } } ({\rm MeV})$ & ${ { E }_{ \text{R} } } ({\rm MeV})$  \\
				\hline \\
		${ { (1) }^{ 2 } { (2) }^{ 2 } { (3) }^{ 2 } }$ & ${ { 0 }^{ + } }$ & 0  & 0  \\
		${ { (1) }^{ 2 } { (2) }^{ 2 } (3)^1(3)^1 }$ & ${ { 2 }^{ + }, { 4 }^{ + } }$ & 5.788  & 5.328  \\
		${ { (1) }^{ 2 } { (3) }^{ 2 } (2)^1(3)^1 }$ & ${ { 2 }^{ + }, { 3 }^{ + } }$ & 5.730  & 5.459  \\
		${ { (1) }^{ 2 } { (3) }^{ 4 } }$ & ${ { 0 }^{ + } }$ & -- & -- \\
		${ { (2) }^{ 2 } { (3) }^{ 2 } (1)^1(3)^1 }$ & ${ { 2 }^{ - }, { 3 }^{ - } }$ & 7.668  & 7.820  \\
		${ { (3) }^{ 4 } (1)^1(2)^1 }$ & ${ { 0 }^{ - }, { 1 }^{ - } }$ & 7.744  & 8.096  \\
		${ { (2) }^{ 2 } { (3) }^{ 4 } }$ & ${ { 0 }^{ + } }$ & 9.161  & 9.846  \\
		${ { (1) }^{ 2 } { (3) }^{ 2 } (3)^1(3)^1 }$ & ${ { 2 }^{ + }, { 4 }^{ + } }$ & 9.166  & 8.449  \\
		${ { (2) }^{ 2 } { (3) }^{ 2 } (3)^1(3)^1 }$ & ${ { 2 }^{ + }, { 4 }^{ + } }$ & 14.041  & 14.059  \\
	\end{tabular}
	\caption{The initial configuration ($G=0$) and energies of different states of ${ {  }^{ 18 } }$C calculated using both the generalized Richardson equations ($E_{\rm GR}$) and the standard Richardson equations ($E_{\rm R}$). The second ${ { 0 }^{ + }_2 }$ state could not be calculated due to a singularity problem arising at a finite ${ G }$. For other details, see the caption of Fig. \ref{carbon14_conf}.}
	\label{carbon18_conf}
	\end{table}	
	\begin{figure}[ht]
		\centering
		\includegraphics[width=0.5\linewidth]{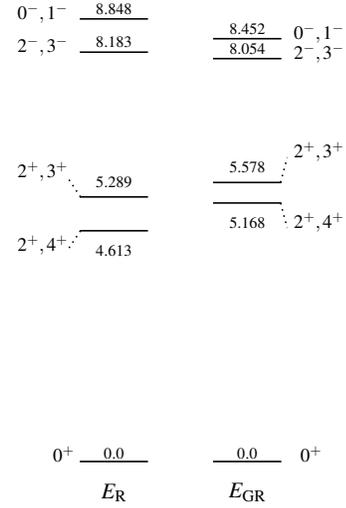}
		\caption{The spectrum of ${ {  }^{ 20 } }$C calculated using the standard Richardson equations (no continuum) $(E_{\rm R}$) is compared with the spectrum obtained by solving the generalized Richardson equations ($E_{\rm R}$). For more details, see the discussion in the text. No excited states are known experimentally for this nucleus.			}
	\label{fig_20C_spectrum}
	\end{figure}
	\begin{table}[h!]
	\centering
	\begin{tabular}{ccccc}
		Config & State & ${ { E }_{ \text{GR} } }$ ({\rm MeV})& ${ { E }_{ \text{R} } }$ ({\rm MeV}) \\
		\hline \\
		${ { (1) }^{ 2 } { (2) }^{ 2 } { (3) }^{ 4 } }$ & ${ { 0 }^{ + } }$ & 0 & 0 \\
		${ { (1) }^{ 2 } { (2) }^{ 2 } { (3) }^{ 2 } (3)^1 (3)^1 }$ & ${ { 2 }^{ + } , { 4 }^{ + } }$ & 5.168 & 4.613 \\
		${ { (1) }^{ 2 } { (3) }^{ 4 } (2)^1(3)^1 }$ & ${ { 2 }^{ + }, { 3 }^{ + } }$ & 5.578 & 5.289 \\
		${ { (2) }^{ 2 } { (3) }^{ 4 } (1)^1(3)^1 }$ & ${ { 2 }^{ - }, { 3 }^{ - } }$ & 8.054 & 8.183 \\
		${ { (3) }^{ 6 } (1)^1(2)^1 }$ & ${ { 0 }^{ - }, { 1 }^{ - } }$ & 8.452 & 8.848 \\
	\end{tabular}
	\caption{The initial configuration ($G=0$) and energies of different states of ${ {  }^{ 20 } }$C calculated using both the generalized Richardson equations ($E_{\rm GR}$) and the standard Richardson ($E_{\rm R}$) equations is compared with the experimental spectrum. We omitted configurations with more than 2 pairs on a level.}
	\label{carbon20_conf}
	\end{table}	

	Fig. \ref{fig_14C_spectrum} presents the spectrum of $^{14}$C calculated using either the generalized Richardson equations for the rational Gaudin model with the continuum, or the standard Richardson equations for the same model but without the continuum. The experimental spectrum for this nucleus is shown for a comparison. The pairing strength in both calculations is adjusted to reproduce the experimental ground state energy of $^{14}$C with respect to $^{12}$C. The calculated spectra in both models are identical, except for the excited $0^+$ states which are shifted down by the coupling to the continuum. The first excited $0^+$ state is shifted by almost 400 keV with respect to the ground state even though the experimental one- and two-neutron separation energies in this nucleus are large. Identical energy for other states is an artifact of having $^{12}$C as a core, namely, these states can be created only by breaking a pair of valence neutrons in $^{14}$C.  The pairing correlations in this case are absent and so are the continuum effects. For each calculated state of $^{14}$C, initial configurations and excitation energies are shown in Table \ref{carbon20_conf}. The initial configuration ($G$=0) is defined by an index of an occupied level, {\em e.g.} $1\equiv 0p_{1/2}, 2\equiv 1s_{1/2}, 3\equiv 0d_{5/2}$, etc. and the number of particles in a given level ($n=1, 2,\dots$). $n=1$ means an unpaired particle. $n=2$ or 4, denotes 1 or 2 pairs of particles, respectively.
	
	Fig. \ref{fig_16C_spectrum} presents the spectrum of $^{16}$C. Both the generalized Richardson equations and the standard Richardson equations for the same model without the continuum fail to reproduce an experimental sequence of states. This is a failure of the schematic two-body interaction in this model. Comparing the spectra of $^{16}$C obtained in the two variants of the rational Gaudin model, one may notice significant relative energy shifts which depend strongly on the configuration of a given state. The individual shifts due to the continuum couplings in this model can be as large as 600 keV. Similar conclusions can be made by comparing results of the rational Gaudin model, with and without the continuum couplings, for $^{18}$C
	(Fig.  \ref{fig_18C_spectrum}) and $^{20}$C (Fig. \ref{fig_20C_spectrum}).
	
	These examples show that the continuum couplings in the rational Gaudin model have significant and non-trivial effect on the spectra of studied systems. Adjusting parameters of the SM Hamiltonian in one nucleus, $^{14}$C in the studied chain of isotopes, to include effectively neglected continuum effects does not solve the problem in heavier isotopes of the same chain for which significant state and configuration dependent energy shifts due to the continuum couplings are found.
	
	Even though the rational Gaudin model is not a realistic approximation of nuclear SM Hamiltonian, one is tempted to conclude that results are more general than the model itself, {\em i.e.} the coupling between discrete and continuum states cannot be replaced by simply fitting the two-body matrix elements to the observed spectra in a certain mass region. This standard procedure in many practical applications leads to wrong conclusions about the nature of effective interactions and the structure of many-body states. This is particularly worrisome if one wants to study states in long chains of isotopes from the valley of stability towards the drip lines.
	
	}

\section{Summary and conclusion}
\label{sec4}
{
Algebraic models, based on emergent symmetries of nuclear many-body problem, helped in the past to identify elementary building blocks and essential concepts behind the formation mechanism of rich spectra of excited states. In the domain of weakly bound and/or unbound nuclei, such models do not exist, what hinders the understanding of qualitative features of the continuum in nuclear spectroscopy.  The pairing model plays a special role among the algebraic models. Exact solution for this Hamiltonian was derived by Richardson for a spectrum of bound s.p. levels \cite{richardson63_1,richardson64_2}. 

In this work, the pairing Hamiltonain was extended to Berggren basis and the generalized Richardson solution was derived for this problem. The comparison between this solution and exact results of GSM, obtained by the diagonalization of the pairing Hamiltonian, confirmed that the generalized Richardson solution is a reliable alternative of an exact GSM diagonalization, in particular in heavy nuclei with large number of valence nucleons.

The chain of carbon isotopes was studied using the generalized Richardson solution for a schematic pairing Hamiltonian in two approximations: (i) in the closed quantum system approximation, {\em i.e.} with bound s.p. levels and neglecting continuum couplings, and (ii) in the open quantum system approximation using Berggren s.p. ensemble.
Fixing in both approaches the strength of pairing interaction in a nucleus ($^{14}$C) with 2 nucleons outside the $^{12}$C closed core, it was found that the $A$-dependence of binding energy and the spectra of $^{14-20}$C rely on the continuum coupling. Another observation was that the effect of continuum coupling on eigenvalues and eigenfunctions, depends strongly on the coupling of nucleons and, hence, varies rapidly from one state to another.
Of course, the interaction in this model is too simple but the qualitative effects are indisputable. They give a warning that results of SM should be interpreted with caution as this model could miss significant physical ingredients.

In the problem of ultra-small superconducting grains, the generalized Richardson solution of the pairing Hamiltonian in Berggren basis could help to understand the influence of continuum on pairing properties, in particular in the transitional region of the weak coupling limit. At present, the lack of experimental data hinders the application of the generalized Richardson equations.

}

\begin{acknowledgements}
We thank R. Id Betan for useful discussions. J.D. is supported by grant No. FIS2015-63770-P (MINECO/FEDER), and N.M. is supported
by the U.S. Department of Energy, Office of Science, Office of Nuclear Physics under award numbers DE-SC0013365 (Michigan State University).
\end{acknowledgements}

\bibliographystyle{apsrev4-1}

\end{document}